\documentclass[aps,prb,showpacs,groupedaddress,twocolumn]{revtex4}
\usepackage{graphicx,bm}
\begin{document}
\title{A macroscopic origin of the Hall anomaly in the mixed state of
type-II superconductors}
\author{P. Tekiel }
\affiliation{Institute of Low Temperature and Structure Research, Polish Academy of
Sciences, 50-950 Wroc\l {}aw, P.O.Box 1410, Poland}
\begin{abstract}
The Hall voltage in the mixed state close to the superconducting critical
temperature is determined in the framework of the macroscopic approach. Only
flux flow and macroscopic excitations motion are taken into account. The
Hall anomaly can appear for rather low magnetic field values.
\end{abstract}

\maketitle

\ \ The Hall effect in the mixed state of many type-II superconductors has a
puzzling feature. Namely, in most of cuprates and some conventional
superconductors the Hall voltage changes its sign as the materials enter the
superconducting state. The classical theories of vortex motion as Bardeen-
Stephen and Nozieres- Vinen models\cite{1,2} seem to be inadequate to
explain this effect. There exist many different attempts to explain this
curious phenomenon. The theory based on the time dependent Ginzburg- Landau
equation\cite{3,4} shows that the Hall conductivity in the flux flow state
can be expressed as the sum of the contribution due to the quasiparticles
and the Hall term due to the vortex flow. If these contributions have
opposite signs then the anomaly can take place. It depend on the electronic
structure of the material. In other models the anomaly is ascribed to the
pinning force~\cite{5}, fluctuation effects~\cite{6,7,8} or due to the
difference of the electron density between the center and outside region of
vortices.~\cite{9,10,11} Our aim is to show that the macroscopic excitation
(ME) motion gives an important contribution to the sign reversal of the Hall
voltage. As it will be seen in the following the sign of the Hall voltage
produced by the ME motion is always opposite to that one due to the flux
flow (or quasiparticles). When the ME motion gives the sufficiently large
rise to the measured voltage then the anomaly occurs.

The macroscopic approach~\cite{12,13,14,15} describes the mixed state in the
intermediate field region $(H_{c1}<<H_{e}<<H_{c2})$, where $H_{c1}$and $%
H_{c2}$ is the lower and upper critical field, respectively. $H_{e}$ is the
external magnetic field intensity. In this description the mixed state
appears as a continuous medium. The magnetic induction vector $\mathbf{B}$
and macroscopic current density $\mathbf{J=}$ $c\nabla \times \mathbf{B/}%
4\pi $ are averaged over the volume with dimension larger than the London
penetration depth $\lambda $. The characteristic length of the macroscopic
current penetration $\delta =\lambda \left( B/4\pi M\right) ^{1/2}$, where $%
4\pi M\approx \Phi _{0}\left( 8\pi \lambda ^{2}\right) ^{-1}\ln \left(
H_{c2}/B\right) $ and $\Phi _{0}$ is the flux quantum. For the considered
magnetic field range $\left( \lambda ^{2}/\delta ^{2}\right) \equiv p\ll 1$.
In the absence of an external current the homogeneous mixed state with
induction $\mathbf{B}_{0}=const$ is treated as the ground state of the
system. Spontaneous inhomogeneities which minimize the energy of the system
are the macroscopic excitations (MEs).\cite{14} The excitation can have the
form of "vortex" ring or "vortex"-"antivortex" pair (in thin slab). The
energy of the unit length of ME has the following form \cite{14}
\[
E_{M}\approx \left( \frac{p}{2}\right) \left( \frac{\Phi _{0}}{4\pi \delta }%
\right) ^{2}\ln \left( \frac{1}{p}\right)
\]
The excitations carry the magnetic flux quanta, thus take part in the creep
processes, both thermal and quantum.~\cite{14,15} Moreover, MEs motion gives
rise to the electrical resistivity, which is independent on $J$ (Ohm low)
for small current density $J$ in thin samples.~\cite{14} Of course, the
contributions of the MEs to the different processes depend on the density of
the excitations. In the low temperature this amount is exponentially small.
At elevated temperature the resistivity of the mixed state strongly
increases. This effect is usually ascribed to the "melting" of the flux line
lattice (FLL). In sufficiently thin samples placed in a perpendicular
magnetic field this process is treated as the example of the Berezinskii-
Kosterlitz- Thouless (BKT) type transition. Since the energy of the usual
(Abrikosov) vortex \ is large even for thin films the transition is often
considered as dislocation mediated one.~\cite{16,17} The FLL dislocation
energy is considerably less than the usual vortex one, but comparison with
the experiment on NbGe~\cite{18} shows that it is still a few orders of
magnitude too large to explain the experimental data. Let us notice that $%
E_{M}$ is less than the energy of the unit length of the usual vortex by a
factor of $\left( 4\pi M/B\right) ^{2}\equiv p^{2}<<1$. The temperature
dependence of $E_{M}$ in the vicinity of the superconducting critical
temperature $T_{c0}$ is $\left( 1-t\right) ^{3}$, where $t\equiv T/T_{c0}$.
For usual vortex or FLL dislocation the corresponding dependence is $\left(
1-t\right) $. It means that for sufficiently high temperature the
macroscopic "vortex" pairs can mediate BKT- type melting transition even in
the slab of a quite large thickness. In this case BKT transition (crossover)
temperature $T_{M}$ is determined as usual by the following relationship
\[
kT_{M}\simeq gp\left( \frac{\Phi _{0}}{4\pi \delta }\right) ^{2},
\]
where $g$ is the sample thickness and $k$ is the Boltzmann constant. $%
p\equiv \left( 4\pi M/B\right) $. For $\left( 1-t_{M}\right) <<1$, where $%
t_{M}\equiv T_{M}/T_{c0}$ we then obtain the well known relation for the
melting (irreversibility) line:
\[
\left( 1-t_{M}\right) ^{3}\sim B^{2}.
\]

\begin{figure}[!htb]
\includegraphics*[width=0.45\textwidth]{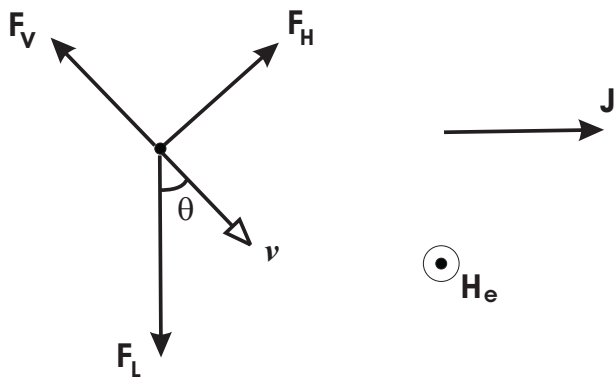}
\caption{Forces acting on the usual, or macroscopic vortex}
\label{Fig1}
\end{figure}

Let us consider a thin slab of type II superconductor placed in the
perpendicular magnetic field $\mathbf{H}_{e}$. An electric current with
density $\mathbf{J}$ is flowing trough the sample paralell to its surface.
In the system there are MEs, both "vortex" rings and "vortex"- "antivortex"
pairs. If $J$ is sufficiently small then the current is able to drive only
the "vortex" pairs.\cite{14} The "vortex" rings are too small and collapse.
If the macroscopic "vortex" is at the point $\mathbf{r}_{i}$ on the slab
then the equation describing the magnetic field distribution connected with
the local mixed state inhomogeneity has the following form~\cite{14}

\begin{equation}
\delta ^{2}\nabla \times \nabla \times \mathbf{b+b=z\Phi }_{0}\delta \left(
\mathbf{r-r}_{i}\right)  \label{eq1}
\end{equation}

where $\mathbf{b=B}\left( \mathbf{r}\right) -\mathbf{B}_{0}$, $\mathbf{B}%
_{0}=const$ and $\delta ^{2}p=\lambda ^{2}$. $\mathbf{z}$ is the unit vector
in $\mathbf{b}$ direction and $\delta \left( \mathbf{r}\right) $ is the two
dimensional Dirac delta. Eq.(1) can be rewritten as

\begin{equation}
\mathbf{j=}\frac{c}{4\pi \delta ^{2}}\left[ \frac{\Phi _{0}}{2\pi }\nabla
\varphi -\mathbf{a}\right]  \label{eq2}
\end{equation}

where $\nabla \times \mathbf{a=b}$ and $4\pi \mathbf{j=}c\nabla \times
\mathbf{b}$. $\varphi $ is the macroscopic phase which gives the flux
quantization. In Eqs (1) and (2) it appears the length $\delta $ instead of $%
\lambda $ of the usual London equation. It means that in the macroscopic
current participates only the small portion $p\rho _{s}$ of the total charge
density $\rho _{s}$ of the superconducting condensate. We assume that under
action of the external current the mixed state in our slab is in a flux flow
state. The temperature is larger than the melting one. So, MEs density is
large and their motion also gives rise to the measured voltage. The forces
acting on the usual vortex and the macroscopic one are of the same kind (see
Fig.~\ref{Fig1}).
$\mathbf{F}_{L}=c^{-1}\Phi _{0}\left( \mathbf{J\times z}\right) $ is the
Lorentz force and $\mathbf{F}_{H}=-\alpha \left( \mathbf{v\times z}\right) $
is the Hall force (or part of the Magnus force). $\mathbf{v}$ is the vortex
(or ME) velocity with respect to the laboratory frame of reference. $\mathbf{%
F}_{v}=-\eta \mathbf{v}$ is the viscous force. Of course, the velocity and
viscosity are different for usual vortex and ME. In a steady state the total
force acting on vortex (or ME) is equal to zero. It can be written as follows

\begin{equation}
\mathbf{F}_{tot}=\mathbf{F}_{L}+\mathbf{F}_{H}+\mathbf{F}_{v}=0  \label{eq3}
\end{equation}

In the almost ideal case the coefficient $\alpha $ for the single vortex in
the Meissner state is equal to $\alpha _{1}=\Phi _{0}\rho _{s}/c$. According
to the above mentioned in the mixed state the charge density that takes part
in the macroscopic current is $p\rho _{s}<<\rho _{s}$. Thus, in the
considered case the coefficient $\alpha =p\alpha _{1}$, both for vortex and
ME. We assume that the viscous drag coefficient $\eta $ for usual vortex is
determined by the Bardeen- Stephen expression
\[
\eta _{f}=\frac{\Phi _{0}H_{c2}\sigma _{n}}{c^{2}},
\]
where $H_{c2}$ is the low temperature value of the upper critical field.\cite%
{1} $\sigma _{n}$ is the normal state conductivity. For ME the corresponding
coefficient has the form~\cite{14,15}

\begin{equation}
\eta _{M}\approx \frac{\Phi _{0}}{4\pi \delta ^{2}B}\eta _{f}.  \label{eq4}
\end{equation}

Let us notice that $\eta _{M}/\eta _{f}$ is very small for $B>>H_{c1}$.
Thus, the ME motion is much faster than the usual vortex one. The forces
acting on macroscopic "vortex" are sketched in Fig.1. Reflecting this figure
with respect to $\mathbf{J}$ -axis we obtain the situation sketch for
"antivortex". The motion of the macroscopic "vortices" and "antivortices"
produces an electric field. The velocity components paralell to the current
direction seem to be the same for "vortex" and "antivortex". It means that
in a rough approximation the MEs motion give no contribution to the Hall
voltage. However, the electric current is connected with the magnetic
induction gradient. The magnetic field free energy associated with the
macroscopic vortex (or antivortex) has the following form~\cite{14,15}

\begin{equation}
f_{1}=\frac{1}{2}\left( \frac{\Phi _{0}}{4\pi \delta }\right) ^{2}
\label{eq5}
\end{equation}

$f_{1}$ is the energy per unit length and it depends on $B$ through\ $\delta
$. It is the origin of an additional force acting on macroscopic vortex and
antivortex. This force has the only one direction (towards larger $B$). The
result is equivalent to the following replacement

\begin{equation}
J\Rightarrow J\left( 1\pm \frac{\Phi _{0}}{8\pi B\delta ^{2}}\right) ,
\label{eq6}
\end{equation}

where minus is for "vortex" case and plus for "antivortex" one. So, the net
Hall voltage due to MEs motion is proportional to the small parameter $%
\left( \Phi _{0}/8\pi \delta ^{2}B\right) $ and it has always the opposite
sign to the corresponding voltage due to flux flow and quasiparticles. We
take into account only two main contributions to the Hall voltage: flux flow
and MEs motion. Denoting by $\theta _{f}$ the Hall angle between the Lorentz
force direction and the usual vortex velocity(flux flow) we have

\begin{equation}
x\equiv \tan \theta _{f}=\frac{p\alpha _{1}}{\eta _{1}}.  \label{eq7}
\end{equation}

As it has been already mentioned the appearence of the factor $p$ follows
from the existence of the characteristic length $\delta $ of the macroscopic
current distribution in the mixed state.If we take into account Eq.(4) then
for Hall angle of ME motion we have

\begin{equation}
\tan \theta _{M}=xD,  \label{eq8}
\end{equation}

where
\[
D=\frac{4\pi \delta ^{2}B}{\Phi _{0}}.
\]
The apparent Hall angle determining the effective Hall voltage due to MEs
motion is obtained from Eqs.(6),(8) and has the following form

\begin{equation}
\tan \theta _{Meff}=-x/2.  \label{eq9}
\end{equation}

The magnitude of the competing Hall voltages depend on densities of usual
vortices (flux flow) and MEs. The usual vortex density is equal to $B/\Phi
_{0}$. With respect to the ME\ density let us notice that in the Meissner
state if $T_{c0}-T<<T-T_{BKT}$, where $T_{BKT}$ is the temperature of BKT-
transition, then the vortex density $n_{f}\simeq \xi ^{-2}$, where $\xi $ is
the superconducting coherence length.~\cite{19} For the macroscopic vortex
in the mixed state $\lambda $ plays the same role as $\xi $ for usual vortex
(cutoff length).~\cite{14} Thus, we may assume that for sufficiently high
temperature the ME density $n_{M}\simeq \lambda ^{-2}$. Now, we can
determine the difference between the Hall voltages (or resistivities) due to
MEs motion and flux flow. Denoting by $\rho _{M}$ and $\rho _{f}$ the Hall
resistivity due to MEs motion and flux flow, respectively, and using
Eqs.(7)-(9) we have

\begin{equation}
\Delta \equiv \left( \rho _{M}-\rho _{f}\right) /\rho _{f}\approx \left[
\frac{\Phi _{0}^{2}}{8\pi \lambda ^{2}\delta ^{2}B^{2}x^{2}}-1\right] ,
\label{eq10}
\end{equation}

where we have assumed $x^{2}<<1$ and $D^{2}x^{2}>>1.$ In the case of YBCO (%
\symbol{126}90K) with $\sigma _{n}\sim 10^{4}\left( \Omega cm\right) ^{-1}$,
$\lambda \sim 10^{-5}cm$ and $\kappa \sim 100$. taking $B\sim 1T$ and $p\sim
10^{-2}$ from Eqs (7)-(8) we obtain $x\sim 10^{-3}$and $D\sim 10^{4}.$If we
assume that only two considered processes contribute to the Hall voltage
then the anomaly is observable for $\Delta >0.$It gives $x<x_{0}$, where $%
x_{0}\approx \Phi _{0}/\left[ 2\left( 2\pi \right) ^{1/2}\lambda \delta B%
\right] $ . For the above quoted YBCO parameters $x_{0}$ has the same order
of magnitude as $x.$ Let us notice that $x_{0}\sim \left( 1-t\right)
^{3/2}B^{-3/2}$ while from Eq.(7) we have $x\sim \left( 1-t\right)
^{2}B^{-1}.$It means that for sufficiently large $B$ or far from $T_{c0}$ we
have $x>x_{0}$ and the ME contribution to Hall anomaly is too small to be
observable. In our case the longitudinal voltage (resistivity) is the sum of
two components:: of flux flow and due to ME motion. They have the same sign.
Each resistivity is proportional to the following expression%
\[
\frac{n_{i}}{\eta _{i}\left( 1+\tan ^{2}\theta _{i}\right) },
\]

where $i=f,M$ denotes flux flow and ME motion, respectively. Putting here
the proper quantities and assuming $x^{2}<<1$ and $D^{2}x^{2}>>1$ we see
that flux flow resistivity is proportional to $B$ and ME one is almost
independent on $B.$ Together with Eqs.(7),(9) it gives the flux flow Hall
voltage independent on $B$ and the effective Hall voltage due to ME motion
proportional to $B^{-1}.$

We have shown that the Hall anomaly has not to follow merely from a
microscopic origin as it is often suggested.~\cite{17,20} We have taken into
account only two competing processes: flux flow and MEs motion. Even in such
simplified situation the Hall anomaly can take place. The quantitative
comparison with experiment should be treated as a rough estimation. The real
materials are usually anisotropic and have often the layered structure.
Moreover, most of experiments have been done on thin films with thickness of
order of $\lambda $ whereas the macroscopic approach used here is derived
for systems with dimensions much larger than $\lambda .$

With respect to the experimental evidence of the macroscopic excitations we
are convinced that the measurements of the Hall effect are little
conclusive, because there are other possible mechanisms which give rise to
the Hall anomaly. Only the experiments which are able to observe the motion
of a single ME can give the conclusive information. As it has been mentioned
the ME\ motion is very fast in comparison with the usual vortex one.
Moreover, the direction of ME motion is almost paralell to the current
direction whereas the usual vortices in the flux flow regime move almost
perpendicularily to this direction.

\end{document}